\documentstyle[12pt,epsf]{article}
\newcommand{\be}{\begin{equation}}
\newcommand{\ee}{\end{equation}}
\newcommand{\ba}{\begin{eqnarray}}
\newcommand{\ea}{\end{eqnarray}}

\begin{document}
\title{Dirac Fermions and Domain Wall Defects
              in $2+1$ Dimensions}
\author{C.D.\ Fosco\thanks{fosco@cab.cnea.edu.ar}\\
and\\
A. L\'opez\thanks{lopez@cabtep2.cnea.edu.ar}
\\
\\
{\normalsize\it
Centro At\'omico Bariloche,
8400 Bariloche, Argentina}}
\maketitle
\begin{abstract}
We investigate some properties of a system of Dirac fermions in $2+1$
dimensions, with a space dependent mass having domain wall like defects.
These defects are defined by the loci of the points where the mass
changes sign. In general, they will be curves lying on the spatial
plane. We show how to treat the dynamics of the fermions in such a way
that the existence of localized fermionic zero modes on the defects is
transparent. Moreover, effects due to the higher, no-zero modes, can be
quantitatively studied. We also consider the relevance of the profile of
the mass near the region where it changes sign. Finally, we apply our
general results to the calculation of the induced fermionic current, in
the linear response approximation, in the presence of an external
electric field and defects.
\end{abstract}
\bigskip
\newpage
\section{Introduction}
In this paper we study some properties of a system of Dirac fermions in
$2+1$ dimensions with a space dependent mass that corresponds to domain
wall like defects. Dirac fermions models in $2+1$ dimensions appear as
effective theories for various systems of interest in particular in condensed matter
physics. Consider for instance a tight binding model for spinless
electrons on a two dimensional square lattice, in the presence of an
external magnetic field of half of a magnetic flux quanta per plaquette.
The nearest neighbor hopping amplitude is one, the second neighbor
hopping amplitude is $t/4$, and there is a staggered potential $\mu(-1)^
{x_1+ x_2} $, where $x_1$ and $x_2$ are the coordinates on the plane. It
has been shown \cite{ludwig} that, in the continuum limit, the effective
theory for this system corresponds to two massive Dirac fermions whose
masses are proportional to $(\mu \pm t )$. In the low energy sector, the
heaviest particle can be neglected and the theory to be considered is a
Dirac fermion with mass proportional to $ M =\mu - t $. Now imagine that
in the system there is a domain wall (or stacking fault) created in the
process of crystal growth by some specially prepared circumstances. At
the level of the effective Dirac theory this can be obtained by
replacing $M$ by $-M$ in half of the plane \cite{boya}. Thus the mass
becomes position dependent with a profile $M(x_1,x_2)= M (\theta(x_2) -
\theta(-x_2))$ for a sharp wall at $x_2=0$. Exactly the same effect can
be obtained when the system, even without any defect, has an odd number
of lattice rows in one direction. If the system is build up periodically
the presence of the extra row produces the same effect as a domain wall.

In this work we study the general case in which the mass of the Dirac
fermions changes sign along an arbitrary curve in the spatial plane, and
with an arbitrary profile near the regions where the sign change is
observed. Due to the Callan and Harvey mechanism \cite{callan}, 
localized chiral zero modes appear on the defects. 
This property has been extensively applied, following an idea by 
Kaplan~\cite{kap1}, to the problem of defining a lattice action for 
chiral fermions, thus providing a possible way of overcoming the 
kinematical obstruction posed by the Nielsen-Ninomiya 
theorem~\cite{kars}.
The proposal evolved into the somewhat more abstract, but nevertheless
equivalent, `overlap formalism'~\cite{nar,overl,kap2}. In Kaplan's 
formulation, one deals with step-like rectilinear defects for the mass 
in an odd number of spacetime dimensions. 
Based on those ideas and techniques, we shall here study the contribution to the 
effective action of the zero modes appearing along a curve of arbitrary 
shape in the spatial plane in $2+1$ dimensions, quantifying the effect of the 
extended bulk states and of the particular shape of the domain wall 
profile on the dynamics of the modes. Localized non chiral zero modes 
are also taken into account. As an example of an application, we use the 
linear response approximation to derive an expression for the induced 
current in the presence of an external uniform electric field.

The paper is organized as follows: In section 2 we introduce an eigenmode
expansion appropriate to the presence of domain wall defects,
considering first the simplest case of a rectilinear domain wall, and
then extending it to the more general situation of a defect of arbitrary
shape. We apply this to the particular example of a circular defect.
In section 3, the eigenmodes expansion is applied to the problem of
evaluating the vacuum fermionic current in the presence of an external
field, to understand the consequences of having defects to the spatial
distribution of the fermionic current. The results for the concrete example 
of the induced current due to a constant electric field are presented and
discussed. Section 4 is devoted to our conclusions.

\section{Eigenmodes expansions in the presence of domain wall defects}

A functional integral over fermionic (Grassmann) fields usually carries
a quadratic action, in terms of some operator, which depends on
the external fields. They are regarded as sources at this
stage, either because they have no dynamics (as in our case), or because
it would be desirable to postpone their quantisation. The formal
integration over spacetime dependent Grassmann fields may be converted
into one over the (Grassmannian) coefficients for the expansion of the
fields in a complete set of eigenfunctions of some operator. A judicious
choice of eigenfunctions may simplify the task of evaluating the
functional integral, or provide a better starting point for an
approximate treatment. 

In the present case, we deal with Dirac fermions in $2+1$ Euclidean
dimensions, in the presence of an external Abelian gauge field $A_\mu$,
and a space dependent parity breaking mass $M({\vec x})$, with 
${\vec x}=(x_1,x_2)$. The Euclidean action $S$ is defined by 
\be 
S \;=\; \int d^3 x \;{\bar \Psi}(x)\;\left[\not \! \partial + i e \not \!\! A 
+ M({\vec x})\right]\;\Psi (x) 
\label{i01}
\ee 
where $x=({\vec x},x_3)$, $x_3$ being the Euclidean time coordinate. Hermitian 
$\gamma$-matrices are in an irreducible ($2 \times 2$) representation of the 
Euclidean Dirac algebra:
\be 
\{ \gamma_\mu , \gamma_\nu \} \;=\; 2 \, \delta_{\mu \nu} \;. 
\label{i02}
\ee

By definition, the mass $M({\vec x})$ has a domain wall defect whenever
it changes sign. As these defects proceed from the regions where the
surface spanned by the function $M({\vec x})$ crosses the plane 
$M ({\vec x}) = 0$, we see that they shall consist of (possibly disconnected) 
curves. Let 
\be
\tau \;\rightarrow\; {\vec x}(\tau) \;,\;  \tau \in [\tau_1,\tau_2]  
\label{i03}
\ee
be a parametrisation of one of the connected components of the curve.
Then,  of course we shall have
\be
M({\vec x}(\tau))\,=\, 0, \forall \tau \in [\tau_1,\tau_2] \;.
\label{i04}
\ee
This curve can in principle have an arbitrarily complicated structure,
depending of course on the assumptions made about the function $M ({\vec
x})$. We shall assume that the mass is regular enough as to define a
curve which is a differentiable manifold. We will see that the
phenomenon of localization in the general case may be understood in a
simple way, if one defines a convenient system of coordinates in the
neighborhood of each point on the curve. For each coordinate patch, the
physics can be shown to be equivalent to the one of a mass term
depending on only one coordinate. Thus, in each coordinate patch, the
defect will look like rectilinear. Our strategy will be to consider
first the simpler situation of a rectilinear defect, and then to
generalize it to the case of a wall defining an arbitrary curve, through
the use of the above mentioned system of coordinates.

\subsection{Rectilinear defect}
This is indeed the simplest situation, consisting of the mass being a 
function of only one of the spatial coordinates, say 
$x_2$~\footnote{For a step-like defect, this is the $2+1$ analog 
of the configurations relevant to Kaplan's formulation~\cite{kap2}.}, 
and changing sign along the straight line $x_2=0$:
\be
S \;=\; \int d^3 x \;{\bar \Psi}(x) \,{\cal D}\,\Psi (x) 
\label{i11}
\ee 
where we defined 
\be
{\cal D}=\gamma_\mu D_\mu + M (x_2)\;.
\label{i12}
\ee
and $D_\mu = \partial_{\mu} + i e A_{\mu}$.

It is easy to show that in order to be able to disentangle the dynamics 
of the fermions into two pieces: one depending on $x_2$ only, and the 
other one with support in the plane $x_2=0$,
the most general gauge field configuration we can consider is  
\be
F_{2\mu} = 0 \;,\; \mu = 1, 3 \;,
\label{i13}
\ee
which corresponds to having no magnetic field ($F_{21} = 0$), and no
electric field perpendicular to the defect ($F_{23} = E_2 = 0$). These 
are gauge-invariant conditions, which in the gauge $A_2 = 0$ (which we
shall adopt) leads to
\be
\partial_2 A_\mu \;=\; 0 \;\;,\;\;\; \mu \,=\, 1, 3\;.
\label{i14}
\ee
In this gauge, we immediately see that the operator ${\cal D}$ may be rewritten as
follows:
\be
{\cal D} \;=\; (a+\not\!d) {\cal P}_L  + (a^\dagger+\not\!d) {\cal P}_R 
\label{i15}
\ee
where $a, a^\dagger$ are operators acting on functions of $x_2$,
\be
a\;=\;\partial_2 + M(x_2) \;\;,\;\; a^\dagger\;=\;-\partial_2 +
M(x_2) \;,
\label{i16}
\ee
${\cal P}_{L,R}$ are projectors along the eigenspaces of the matrix
$\gamma_2$:
\be
{\cal P}_L \;=\; \frac{1}{2} (1 + \gamma_2) \;\;,\;\; 
{\cal P}_R \;=\; \frac{1}{2} (1 - \gamma_2) \;,
\label{i17}
\ee
and $\not \! d$ is the two dimensional Euclidean Dirac operator
corresponding to the two coordinates $x_1$ and $x_3$, 
which we denote collectively by ${\hat x}$, namely
\be
\not \! d \;=\; \gamma_1 (\partial_1 + i e A_1 ({\hat x}))\,+\, 
\gamma_3 (\partial_3 + i e A_3 ({\hat x})) \;.
\label{i18}
\ee

Expression (\ref{i15}) suggests the possibility of getting rid of the
dependence on $x_2$ for the fields, by a suitable expansion in the modes
of some operator, and obtaining in a way a `dimensional reduction' from
the three dimensional spacetime to the two dimensional one corresponding 
to ${\hat x}$.  
As ${\cal D}$ itself is not Hermitian, we use instead the
positive Hermitian operator ${\cal H}$
\be
{\cal H} \;=\; {\cal D}^\dagger {\cal D} \;,
\label{i19}
\ee
which, by using the explicit expression (\ref{i15}) for ${\cal D}$ leads 
to
\be
{\cal H} \;=\; (h - \not\!d^2) {\cal P}_L + ({\tilde h} - \not \! d^2)
{\cal P}_R \;,
\label{i20}
\ee
with $h=a^\dagger a$ and ${\tilde h}=a a^\dagger$.
The appearance of the two conjugate Hermitian operators $h, {\tilde h}$ suggests
an expansion for the $x_2$ dependence of the fermionic fields in terms of their 
eigenfunctions, namely
\ba
\Psi (x_2,{\hat x}) &=& \sum_n \left[ \phi_n (x_2) \psi^{(n)}_L ({\hat x}) +   
{\tilde \phi}_n (x_2) \psi^{(n)}_R ({\hat x})  \right]\nonumber\\
{\bar \Psi}(x_2,{\hat x}) &=& \sum_n \left[ {\bar \psi}^{(n)}_L ({\hat x}) 
\phi_n^\dagger (x_2) +{\bar \psi}^{(n)}_R ({\hat x}) {\tilde \phi}_n^\dagger (x_2)   
\right]
\label{i21}
\ea
where the subscripts $L, R$ denote the `chirality' defined by the matrix $\gamma_2$   
and 
$$
h \phi_n (x_2) \;=\; \lambda_n^2 \phi_n (x_2)\;\;,\;\;
{\tilde h} {\tilde \phi}_n (x_2) \;=\; \lambda_n^2 {\tilde \phi}_n (x_2)
$$
\be
\langle \phi_n | \phi_m \rangle \;=\; \delta_{n,m} \;\;,\;\;
\langle {\tilde \phi}_n | {\tilde \phi}_m \rangle \;=\; \delta_{n,m}
\;.
\label{i22}
\ee 
To write eq.(\ref{i22}) we have used that 
$\psi^{(n)}_{L,R} ({\hat x})={\cal P}_{L,R} \psi^{(n)}$, and
${\bar \psi}^{(n)}_{L,R} ({\hat x})= {\bar \psi}^{(n)}{\cal P}_{R,L}$.
Note that the two dimensional fermionic fields $\psi^{(n)}$ 
and their Dirac adjoints are independent variables for different values
of the discrete index $n$.
We have made explicit the property that $h$ and ${\tilde h}$ are
positive (the $\lambda_n$ are assumed to be real and we fix their sign,
by convention, to be positive) and have the same spectrum, with the
only possible exception of the zero modes $\lambda_n = 0$, since, for 
any $\phi_n$ with $\lambda_n \neq 0$, there also exists one eigenvector 
of ${\tilde h}$ with identical eigenvalue
\be 
h \phi_n (x_2) \;=\; \lambda_n^2 \phi_n (x_2) 
\;\Rightarrow\; 
{\tilde h} \; [\frac{1}{\lambda_n} a\phi_n (x_2)] \;=\; \lambda_n^2 
\; [\frac{1}{\lambda_n} a\phi_n (x_2)]
\label{i23}
\ee
where the factor $\frac{1}{\lambda_n}$ is introduced to normalize the 
eigenvectors of ${\tilde h}$. (Of course the reciprocal property for 
the eigenstates of ${\tilde h}$ also holds.)
 
We remark that, depending on the specific form of $M(x_2)$, one of the zero 
modes may not appear in the sum (\ref{i21}). In the Callan and Harvey 
mechanism, the defect is step-like, what gives rise to 
only one zero mode, with exponential localisation~\cite{callan}. 
This issue will be made more explicit when considering concrete 
expressions for the mass function. 

We then introduce the expansions (\ref{i21}) into the Euclidean action 
(\ref{i11}), and see that it becomes a sum over actions corresponding to
two dimensional fermions, labeled by the index of the eigenvalue
$\lambda^2_n$
\be
S\;=\; S_L^{(0)} + S_R^{(0)} + \sum_{n\neq 0} \int d^2 {\hat x} 
{\bar \psi}^{(n)}({\hat x}) (\not \! d + \lambda_n) \psi^{(n)} ({\hat x})  
\label{i24}
\ee  
where $S_L^{(0)}$ and $S_R^{(0)}$ are the actions corresponding to the
left and right zero modes, namely
\be
S_L^{(0)}\;=\; \int d^2{\hat x} \;{\bar \psi}_L^{(0)} ({\hat x}) 
\not \! d \psi_L^{(0)} ({\hat x})
\;\;,\;\; 
S_R^{(0)}\;=\; \int d^2{\hat x} \;{\bar \psi}_R^{(0)} ({\hat x}) 
\not \! d \psi_R^{(0)} ({\hat x})
\;.
\label{i25}
\ee

As the fermionic integration measure decomposes into the infinite product:
$$
{\cal D}{\bar \Psi} {\cal D}{\Psi}\;=\;{\cal D}{\bar \psi}_L^0 
{\cal D}{\psi}_L^0 \;{\cal D}{\bar \psi}_R^0 {\cal D}{\psi}_R^0 
$$
\be
\times \prod_{n\neq 0} [{\cal D}{\bar \psi}^{(n)} {\cal D}{\psi}^{(n)}]
\ee
it is evident that the fermionic determinant will be a product of
two dimensional Euclidean determinants, one for each $n$. This yields an 
effective action which decomposes into an infinite sum:
\be
\Gamma \;=\; \Gamma_L^{(0)} + \Gamma_R^{(0)} + \sum_{n\neq 0} \Gamma^{(n)} 
\label{i26}
\ee
where $\Gamma_L^{(0)}$ and $\Gamma_R^{(0)}$ are the effective actions
corresponding to the left and right zero modes, respectively:
\be
\Gamma_L^{(0)}\;=\; -\ln \det (\not \! d {\cal P}_L) \;\;,\;\;
\Gamma_R^{(0)}\;=\; -\ln \det (\not \! d {\cal P}_R) \;,
\label{i27}
\ee
and $\Gamma^{(n)}$ is the effective action for a massive two dimensional 
fermion with mass $\lambda_n$,
\be
\Gamma^{(n)} \;=\; - \ln \det (\not \! d + \lambda_n) \;.
\label{i28}
\ee
Note that the sign of $\lambda_n$, which is chosen by convention, is 
irrelevant to the result of these two dimensional determinants. Namely,
as the spacetime dimension for these modes is even, the mass term does
not break parity and thus the fermionic determinant cannot depend on the
mass sign.

We now discuss, also for the case of a rectilinear defect, the relevance
of the mass profile to the actual properties of the different
contributions to the effective action. Rather than considering defects
of the most general possible kinds, we plan to study domain walls of a
localized type, namely, defects having a typical width $\Delta$. This
width is defined by the extension of the region around a zero of the
fermion mass, where it has an appreciable variation. The mass is assumed
to have an approximately constant absolute value $\Lambda$ outside the
defect. This kind of behaviour is displayed in Figure 1. The analysis
of the different contributions to the fermionic effective action has to
take the finite width of the defect into account. To incorporate the
localized modes, we expand the functional form for the mass around the
defect in a Maclaurin series keeping a few terms, and then solve for
the eigenmodes of $h$ and ${\tilde h}$ corresponding to that expansion.
This procedure will only make sense if the modes so obtained are
concentrated in a region contained in the band of width $\Delta$. So we
shall only add in the sum over modes (\ref{i26}) those having
a spatial dispersion smaller than $\Delta$. This sum will give an
approximate estimation of the contribution of the localized modes to the
effective action. Non-localized modes correspond to Dirac fermions that
live effectively in $2+1$ dimensions, and shall have a smaller
contribution to the effective action than the localized modes, as a
simple argument shows: The action for a non-localized mode 
is given by the corresponding term in the sum (\ref{i24}). The property
that distinguishes it from the action for a localized mode is that 
the former corresponds to an unbounded state of the operators $h$, 
${\tilde h}$, while the latter is due to a bounded state.
These operators can be written more explicitly as,
\ba
h &=&-\partial^2 + M^2 (x_2) -\partial_2 M(x_2) \nonumber\\
{\tilde h} &=& -\partial^2 + M^2 (x_2) + \partial_2 M (x_2)
\ea
which, far from the defect (for a defect like the one of Fig.1) 
may be replaced by
\be
h \;\simeq\; {\tilde h} \;\simeq\; -\partial^2 + \Lambda^2 \;,
\ee 
since the derivative of the mass obviously tends to zero.
We conclude that non-localized modes have an action, which has to be larger or 
equal than $\Lambda$, the absolute value of the mass outside the defect.

If the mass function is approximately linear near to the center of the
defect, we may keep only the first term in a Maclaurin expansion,
arriving to the operators $a$ and $a^\dagger$
\ba
a &=& \partial_2 + M(x_2)\;=\;\partial_2 + M'(0) \,x_2 \nonumber\\
a^\dagger &=& -\partial_2 + M(x_2)\;=\;-\partial_2 + M'(0) \,x_2
\ea
in terms of which we may define ${\hat a}$ and ${\hat a}^\dagger$, which
are identical to the usual (harmonic oscillator) creation and destruction 
operators:
$${\hat a} \;=\; \frac{1}{\sqrt{2 M'(0)}} \; a \;\;, 
\;\;{\hat a}^\dagger \;=\; \frac{1}{\sqrt{2 M'(0)}} \; a^\dagger$$
\be
\;\;\;\Rightarrow\; [{\hat a},{\hat a}^\dagger] \;=\; 1 \;. 
\ee
We note that, as far as this linearization is reliable, we may of course
write $M'(0)$ in terms of $\Lambda$ and $\Delta$:
\be
M'(0)\;=\;\frac{2\Lambda}{\Delta} \;.
\ee 

The eigenvalues and eigenvectors of $h$ and ${\tilde h}$ are immediately 
found. We see that there is a zero mode for ${\hat a}$ but none for 
${\hat a}^\dagger$, 
\be
{\hat a} \phi_0 (x_2) \,=\, 0 \;\Rightarrow\; 
\phi_0 (x_2) \,=\, N_0 \, e^{-\frac{1}{2} |M'(0)| x_2^2}
\ee
with $N_0$ a normalisation constant. As the Hamiltonians $h$ and ${\tilde h}$
are like number operators, the eigenvalues $\lambda_n \;=\; \sqrt{2|M'(0)| n}$ 
are evenly spaced. Then the effective action becomes
\be
\Gamma\;=\; \Gamma_L^{(0)} \,+\, \sum_{n=1}^\infty \Gamma^{(n)}
\ee
where
\be
\Gamma^{(n)}\;=\;{\rm Tr} \ln \left[ \not \! d + \sqrt{2 |M'(0)| n} \right] \;.
\ee
In order to determine the number $N$ of modes that we may include in the sum
above, we recall that for the harmonic oscillator states corresponding
to the label $n$, the dispersion $\sigma_x^{(n)}$ is
\be
\sigma_x^{(n)} \;=\; \sqrt{\frac{n+\frac{1}{2}}{M'(0)}} \;. 
\ee
The condition that modes are localized inside the defect becomes
$\sigma_x^{(N)}=\Delta$, so
\be
N \;=\; [ M'(0) \Delta^2 - \frac{1}{2} ]  \;.
\ee
As $N \geq 0$, we see that 
\be
M'(0) \Delta^2 \geq  \frac{1}{2}
\ee
or
\be
\Lambda \times \Delta \; \geq \; \frac{1}{2}\;,
\ee
which must be interpreted as a sufficient condition to fulfill in order
to have at least one localized state. Its physical interpretation is
quite transparent: the thinner the defect, the bigger should the mass
jump be in order to localize a mode inside the defect.

We can then exhibit the contribution to the effective action of these
modes. The zero mode contribution is simply the effective action for a
chiral fermion in $1+1$ dimensions, a result we recall, for example, 
from reference~\cite{cdet},
\be
\Gamma_L \;=\; 
\frac{e^2}{8\pi} \int d^2x  A_\mu \left[ a \delta_{\mu\nu} - 
(\delta_{\mu\alpha}+i\epsilon_{\mu\alpha}) 
\frac{\partial_\alpha\partial_\beta}{\partial^2} 
(\delta_{\beta\nu}-i\epsilon_{\beta\nu})\right] A_\nu \;.
\ee
where $a$ is a regularization parameter, which must be fixed to the
value $1$ for the regularization to be gauge invariant.

Regarding the massive modes contributions, we apply the result presented 
in the Appendix for the massive  determinant in the quadratic approximation, 
with a mass given by $\lambda_n$, to obtain the $n$ mode contribution:
\be
\Gamma^{(n)}(A)\;=\; \frac{1}{2} \int \frac{d^2 k}{(2 \pi)^2}
{\tilde A}_\mu (-k) {\tilde \Gamma}_{\mu\nu}^{(n)} (k) {\tilde A}_\nu (k)
\ee
where
\be
{\tilde \Gamma}_{\mu\nu}^{(n)} (k)\;=\; {\tilde \Gamma}^{(n)}(k) 
(\delta_{\mu\nu}-\frac{k_\mu k_\nu}{k^2}) \;,
\ee
and
\be
{\tilde \Gamma}^{(n)}(k) \;=\; \frac{e^2}{\pi} \left\{1 - 2 \frac{\lambda_n^2}{k^2} 
(1+\frac{4\lambda_n^2}{k^2})^{-\frac{1}{2}} 
\ln \left[\frac{(1+\frac{4\lambda_n^2}{k^2})^{\frac{1}{2}} + 
1}{(1+\frac{4\lambda_n^2}{k^2})^{\frac{1}{2}} - 1}               
\right] \right\}   
\ee
Then the effective action reads
$$
\Gamma (A) \;=\; \Gamma_L (A) \,+\, 
$$
\be
\frac{1}{2} \int \frac{d^2 k}{(2 \pi)^2}
{\tilde A}_\mu (-k) {\tilde \Gamma}_{v}(k)
(\delta_{\mu\nu}-\frac{k_\mu k_\nu}{k^2}) 
{\tilde A}_\nu (k)
\ee
where
\be
{\tilde \Gamma}_{v}(k)\;=\; \sum_n {\tilde \Gamma}^{(n)}(k)\;
\ee
If a low momentum expansion for ${\tilde \Gamma}^{(n)}$ is used, then we 
have for ${\tilde \Gamma}_v$ the following expression:
\be
\Gamma_v (k)  \;=\; \sum_{n}\Gamma^{(n)}(k) \;=\; 
\frac{e^2}{12 \pi |M'(0)|}\sum_{n=1}^{M'(0) \Delta^2 - 
\frac{1}{2}} \frac{1}{n} \; k^2 \;.
\ee
Moreover, for cases where the condition $\Lambda \Delta >> 1$ (the number of
localized modes is $>> 1$) is satisfied, 
the sum over $n$ can be approximated by a logarithm, leading to
\be
\Gamma_v \;\simeq\; \frac{e^2}{12 \pi |M'(0)|}[\ln (\Lambda \Delta) +
\gamma] k^2 \;,
\ee 
where $\gamma$ is Euler's constant.

In order to understand the nature of the error involved in the assumption 
of taking a linear profile for the mass near the defect, we recall the
the expression for the error in a Maclaurin expansion to first order
may be written as
\be
E (x_2) \;=\; \frac{1}{2} M''(\theta) x_2^2
\ee
where $0 \leq \theta \leq x_2$, and we assumed that $M$ has a continuous
second derivative. A rough estimate of the value $\delta$ of the radius
of the interval in which we may use the linear approximation is
by imposing the condition of having an error much smaller than the value 
of the linear term, which leads to
\be
\delta \;<<\; 2 \frac{|M'(0)|}{|M''({\hat \theta})|} \;,
\ee
where $M''$ reaches its maximum at ${\hat \theta}$.

We have considered so far only the case of one single defect. To see
how the results for this case may be extended to the situation of
having more than one defect, we note that our definition of localized
modes assures that modes attached to different defects shall be,
to a very good approximation, orthogonal to each other. Then the
effective action will be the sum of the effective actions corresponding
to the modes in the different defects. This  shows {\em a posteriori\/}
the utility of the truncation of the sum over modes according to their
dispersion: it ensures the additivity of the corresponding contributions
to the action.

\subsection{Defects of arbitrary shape}
We shall consider here the case of a wall having the shape of an arbitrary 
curve lying on the plane.
For a curve ${\cal C}$ defined by a parametric equation (\ref{i03}), we introduce 
the (normalized) tangent, normal and binormal vectors, denoted by ${\vec e}_1$, 
${\vec e}_2$ and ${\vec e}_3$, respectively~\footnote{Of course, being the curve
plane, it is torsionless (${\vec e}_3$ is constant), or, equivalently,
the osculating plane coincides with the plane $x_3\,=\,0$ for all the points on the
curve.}.
To that end, it is better to use a natural parametrisation for the curve, defined 
in terms of the arc length parameter $s$
\be
s(\tau) = \int_{\tau_1}^\tau \, d\tau'\, 
\sqrt{\frac{d{\vec x}(\tau')}{d\tau'}\cdot\frac{d{\vec x}(\tau')}{d\tau'}}
\ee
\be
s \;\rightarrow\; {\vec x}(s) \;,\;  s \in [0,L] 
\ee
where $L$ is the total length of the curve.
The tangent ${\vec e}_1$ is then obtained as a derivative of the parametric 
equations of the curve with respect to the length parameter 
\be
{\vec e}_1 \;=\; \frac{d}{ds} {\vec x}(s) \;.
\ee
It is natural to define an orientation for the curve: we shall assume
that $s$ increases when the region of positive mass is left to the right
and the one of negative mass is to the left of the curve.
The normal can, as usual, be defined by
\be
{\vec e}_2 \;=\; \pm ||\frac{d{\vec e}_1(s)}{ds} ||^{-1} 
\frac{d{\vec e}_1(s)}{ds}
\ee
with a choice of sign which we fix by the convention that the normal
should point from the side of positive to the one of negative mass.
As the curve is plane, however, we can do more simply than that and
define the normal as follows:
\be
e_2^j \;=\; \epsilon^{jk} e_1^k
\ee
which is explicitly orthogonal to ${\vec e}_1$. It is immediate to
realize that ${\vec e}_3$ can be taken as a constant and pointing in the
third direction, with a sense such that $\det e^i_j \,=\, +1$.
We now proceed to define coordinates in a neighborhood of the defect. 
Through each point on ${\cal C}$, we draw a curve tangent to the
normal vector, and another one tangent to the binormal. Of course the 
condition of being tangent only defines the curves in a small neighborhood, 
but this is enough for our purposes. We then perform a Lie dragging of
the curve ${\cal C}$ along those fields. This defines a complete set of
coordinate lines around the curve of the defect. 

Denoting by $u_1$, $u_2$ and $u_3$ the parameters labeling points along
each one of the integral lines, we may then write the action corresponding
to the region around the defect as
\be
S_{\cal C}\;=\; \int d^3 u \, \det[e^i_j] {\bar \Psi} 
\left[ \gamma^\alpha e^i_\alpha D_i + M (u_1) \right] \Psi 
\ee 
where the spin connection term vanishes because the coordinate system
has been defined by dragging. 
We then note that $\det[e^i_j]=+1$, that $e^i_\alpha \partial_i$ is the 
directional derivative along the integral curve $\alpha$, and that 
$e^i_\alpha A_i$ is the gauge field component along the same direction.
In order to arrive to a situation entirely similar to the one of the
case of the rectilinear defect, we need to chose a convenient gauge. 
The analog of the condition $A_2=0$ for that case would be here to use a
gauge such that the component of  $A$ along the normal to the curve
vanishes. We note that a condition like that does not forbid the
presence of a non vanishing magnetic flux through the surface enclosed by
the curve, since that requires only the existence of a non vanishing tangent 
component for the gauge field. We then arrive to the expression for the action
\be
S_{\cal C}\;=\; \int d^3 u \, {\bar \Psi} 
\left[ \gamma^1 D_{u_1} + \gamma^2 \partial_{u_2} +
\gamma^3  D_{u_3} + M (u_2) \right] \Psi \;.
\label{scd}
\ee

We conclude this section with an example of an application:
the case of a circular defect. We assume the defect is a circle of 
radius $R$, and that the mass is positive inside the circle. The 
convenient choice of coordinates for this case corresponds to using
polar coordinates.
 
With the natural parametrisation, in terms of the arc length $s$, 
we have:
\be
{\vec x}(s)\;=\; R (\; \cos(\frac{s}{R}) , \sin(\frac{s}{R}) \;) 
\;\;\;,\;0 \leq s < 2 \pi R\;.
\ee
The tangent and normal vectors are, of course,
\ba
{\vec e}_1 &=& (-\sin(\frac{s}{R}) , \cos(\frac{s}{R}) ) \nonumber\\
{\vec e}_2 &=& (\cos(\frac{s}{R}) , \sin(\frac{s}{R}) ) \;.
\ea
Local coordinates are defined as follows: $u_1 = r \theta$, 
where $r$ is the radial (plane) distance to the origin and $\theta$ the 
polar angle (of course, $u_1$ reduces to $s$ when $r=R$). $u_2$ is
equal to $r-R$ and $u_3=x_3$. In terms of $r$, $\theta$ and $x_3$,
the action (\ref{scd}) for a defect of width $\Delta$ becomes
\be
S_{\cal C}=\int_{R-\Delta}^{R+\Delta} dr r  
\int_0^{2\pi} d\theta \int dx_3 
{\bar \Psi} \left[ \frac{1}{r} \gamma^1 
(\partial_\theta + i e A_\theta) + \gamma^2 \partial_r +
\gamma^3  D_{x_3} + M (r-R) \right] \Psi \,.
\ee

We may now consider, for example, a gauge field which corresponds
to a static uniform magnetic field $B({\vec x})$ through the surface.
Near to the defect, we may take
\be
A_\theta \;=\; \frac{\Phi}{2 \pi R}
\ee  
where $\Phi$ is the total flux $\Phi = B \pi R^2$. Note that, as we are
only considering the fields near to the defect, the dependence of 
$A_\theta$ on the radius may be ignored, evaluating it at $R$. This will
hold true as long as the radius of the defect is sufficiently large. 
The action for the chiral zero mode localized on the defect will then
be
\be
S_L \;=\; R \int_0^{2 \pi} \, d\theta \,\int dx_3 \,
 {\bar \psi}^{(0)}_L
( \frac{1}{R} \gamma^1 (\partial_\theta + i e \frac{\Phi}{2 \pi R})
+ \gamma_3 \partial_3 )
\psi^{(0)}_L 
\ee
which corresponds to chiral fermions in the presence of a constant
gauge field. This problem is equivalent to the one of free fermions
with twisted boundary conditions, and the dynamics of the zero mode
is then periodic in the flux.
Non zero modes might also be incorporated, but only if their dispersion
is small enough as to  ignore de dependence of the gauge field on the
radial direction.

\section{Fermionic current density}
In this section we present an evaluation of the expectation  value of
the fermionic current in the presence of defects, and under the
influence of an external electric field. From what we said in the
previous section, it should become evident that what we need is just
to study the case of one single defect, and then the more general
situations are easily understood from the knowledge of the results that
follow for this case. 
We shall consider the vacuum expectation value $j_\mu (x_1,x_2,x_3)$
($\mu=1, 3$) of the fermionic current in the presence of an external 
gauge field such that $F_{2\mu}=0$. That expectation value may of course 
be written as 
\be
j_\mu (x) \;=\; - e {\rm tr} \left[ \gamma_\mu S(x,x) \right] 
\ee
where $S(x,y)$ is the fermionic propagator, the inverse of the operator
defining the quadratic form in the fermionic action:
\be
S_{\alpha\beta}(x,y)\;=\;\langle x,\alpha|{\cal D}^{-1}|y,\beta\rangle 
\ee
where
\be
{\cal D} \;=\; (a+\not\!d) {\cal P}_L + (a^\dagger+\not \! d) {\cal P}_R
\;.
\ee
The simple algebraic identity
\be
{\cal D}^{-1} \;=\;  ({\cal D}^\dagger {\cal D})^{-1} {\cal D}^\dagger
\ee 
leads to
\be
{\cal D}^{-1} \;=\;(h-\not\!d^2)^{-1} {\cal P}_L (a^\dagger-\not\!d)
+ ({\tilde h}-\not\!d^2)^{-1} {\cal P}_R (a -\not\!d) \;.
\ee
The $x_2$ dependence of the propagator is easily extracted by using the
completeness of the eigenfunctions of $h$ and ${\tilde h}$, obtaining
$$
S_{\alpha\beta}(x,y)\;=\;\sum_{n=0,m=1}^{\infty}
\phi_n (x_2) {\tilde \phi}^\dagger_m (y_2)
\langle x_1,x_3,\alpha|(\lambda_n^2-\not\!d^2)^{-1}\lambda_n {\cal P}_L
|y_1,y_3,\beta\rangle \delta_{n m}
$$
$$
-\sum_{n=0,m=0}^{\infty}
\phi_n (x_2) \phi^\dagger_m (y_2)
\langle x_1,x_3,\alpha|(\lambda_n^2-\not\!d^2)^{-1} \not\!d {\cal P}_R
|y_1,y_3,\beta\rangle \delta_{n m}
$$
$$
+\sum_{n=1,m=0}^{\infty}
{\tilde \phi}_n (x_2) \phi^\dagger_m (y_2)
\langle x_1,x_3,\alpha|(\lambda_n^2-\not\!d^2)^{-1} \lambda_n {\cal P}_R
|y_1,y_3,\beta\rangle \delta_{n m}
$$
\be
-\sum_{n=1,m=0}^{\infty}
{\tilde \phi}_n (x_2) {\tilde \phi}^\dagger_m (y_2)
\langle x_1,x_3,\alpha|(\lambda_n^2-\not\!d^2)^{-1} \not\!d {\cal P}_L
|y_1,y_3,\beta\rangle \delta_{n m} \;.
\ee
Whence, the expectation value of the current becomes
$$
j_\mu (x) \;=\; -e \phi_0^2(x_2) {\rm tr} [ \gamma_\mu 
\langle x_1,x_3| \not\!d^{-1} {\cal P}_R | x_1,x_3 \rangle ]
$$
$$
+ e \sum_{n=1}^\infty \left\{
\phi_n^2 (x_2) {\rm tr}[\gamma_\mu
\langle x_1,x_3|(\lambda_n^2-\not\!d^2)^{-1} \not\!d {\cal P}_R
|x_1,x_3\rangle] \right.
$$
$$
+{\tilde \phi}_n^2 (x_2) {\rm tr}[\gamma_\mu
\langle x_1,x_3|(\lambda_n^2-\not\!d^2)^{-1} \not\!d {\cal P}_L
|x_1,x_3\rangle]
$$
$$
-\phi_n(x_2){\tilde \phi}_n (x_2) {\rm tr}[\gamma_\mu
\langle x_1,x_3|(\lambda_n^2-\not\!d^2)^{-1} \lambda_n {\cal P}_L
|x_1,x_3\rangle]
$$
\be
\left.-{\tilde \phi}_n(x_2) \phi_n (x_2) {\rm tr}[\gamma_\mu
\langle x_1,x_3|(\lambda_n^2-\not\!d^2)^{-1} \lambda_n {\cal P}_R
|x_1,x_3\rangle] \right\}
\ee
The first term on the rhs is the contribution to the current coming from
the chiral mode. Denoting by $j_\mu^L (x)$ and $j_\mu^{(n)}$ the vacuum
expectation values of the two dimensional chiral and vector like
currents, respectively, we see that the above result for the current may
be rewritten as
$$
j_{\mu}(x_1,x_2,x_3)\;=\;\phi_0^2(x_2) j_\mu^L (x_1,x_3)
$$
\be
-\frac{1}{2}\sum_{n=1}^\infty \left[
({\tilde \phi}_n^2(x_2)+\phi_n^2(x_2))\delta_{\mu\nu} 
+ i({\tilde \phi}_n^2(x_2)-\phi_n^2(x_2))\epsilon_{\mu\nu} 
\right] j_\mu^{(n)} \;.
\label{eq:curr}
\ee

We will now apply the result (\ref{eq:curr}) to the determination of the
current due to a constant electric field $E$ along the defects. 
It is evident from (\ref{eq:curr}) that the $x_2$ dependence 
of the current is completely determined by the functions $\phi_n$ and 
${\tilde \phi}_n$. For example, the contribution carried by the chiral zero 
mode shall be concentrated around the defects with a density $\phi_0^2$, 
which is a Gaussian for a linear defect. A step-like defect will in turn 
have an exponential localization. The $x_2$-dependent factors for the 
$n \neq 0$ modes yield contributions which are more extended in $x_2$,
as the dispersion in $x_2$ grows with $n$.

On the other hand, as the external electric field is assumed to be 
constant  in time and uniform in space, there only remains to know 
the $x_1$ dependence,  which of course may appear in the expectation 
values of the two dimensional currents $j_\mu^L$ and  $j_\mu^{(n)}$.
It can be easily checked that, within the linear response 
approximation, the expectation value of every massive current 
$j_\mu^{(n)}$ is zero, due to 
the existence of a gap~\footnote{The conductivity
of a massive fermion in $1+1$ dimensions is proportional to the 
value of ${\tilde \Gamma}(0)$, which vanishes for mass different 
from zero. It is 
however non-zero for a massless fermion.} (the "mass" $\lambda_n$).
The contribution of the chiral zero mode is obtained from the parity-
conserving part of its effective action, and yields exactly one half of
the contribution corresponding to a {\em massless\/} Dirac fermion:
\be
j_1^L (x_1) \;=\; \frac{e^2}{2 \pi} E L  \;,
\ee
where $L$ is the length of the system in the direction $x_1$.

We conclude that the three dimensional current is:
\be
j_1(x_1,x_3)\;=\;\phi_0^2(x_2) \frac{e^2}{2 \pi} E L \;.
\ee
where, as we have already mentioned, the form of $\phi_0^2(x_2) $ 
depends on the particular shape of the defect.

\section{Conclusions}

Motivated by the fact that Dirac fermions with step like masses appear
as effective descriptions of various systems in particular in 
Condensed Matter Physics, in this paper  we have provided a formalism 
to study the general case in which the mass of the Dirac 
fermions changes sign along an arbitrary curve in the spatial plane, and
with an arbitrary profile near the regions where the sign change is
observed.

 We showed that the dynamics of the fermions can be studied in such 
a way that the contributions from the localized zero modes on the defects, and 
the ones of the higher no-zero modes can be clearly distinguished. 
We provided a criterion 
to evaluate which are the most relevant contributions 
depending on the profile of the mass near the region where it changes sign.

We applied this formalism to the simplest case of a rectilinear domain wall and 
evaluate the vacuum fermionic current in the presence of an external
field, to understand the consequences of having defects to the spatial
distribution of the fermionic current. We studied also a circular defect 
with a magnetic field piercing the enclosed surface and saw that the effective theory for 
the zero mode is equivalent to the one of fermions with twisted boundary conditions in a 
circle.

\newpage
\section*{Appendix: Two-dimensional massive fermionic determinant in the
quadratic approximation}
We present here, for the sake of completeness, the calculation of the
massive fermionic determinant in two dimensions, in the presence of an
external Abelian vector field~\cite{cole} $A_\mu$, in the quadratic 
approximation (namely, the first non-trivial term in the usual 
perturbation expansion).

The generating functionals ${\cal Z}(A)$ and $\Gamma (A)$, corresponding
to massive Dirac fermions in the presence of an external Abelian vector
field $A_\mu$ are defined by
$${\cal Z}(A)\,=\,\exp [-\Gamma (A)] \;=\; \int {\cal D}{\bar \psi}
{\cal D}\psi \, \exp\{-\int d^2 x {\bar \psi} (\not\!\partial + i e \not
\! A + m) \psi \}$$
\be
\,=\,
\det (\not\!\partial + i e \not \!\! A + m) \;.
\label{a01}
\ee
Thus the effective action $\Gamma$ is then given by 
\be
\Gamma (A) \,=\, - \ln \det (\not \! \partial + i e \not \! A + m)
\,=\, - {\rm Tr} \ln (\not \! \partial + i e \not \! A + m)\,,
\label{a02}
\ee
where the trace operates over both functional and Dirac space. In the
quadratic approximation, $\Gamma$ becomes 
\be
\Gamma (A) \;=\; - \frac{1}{2} e^2 {\rm Tr} 
\left[(\not \! \partial + m)^{-1} \not \!\! A\, (\not \! \partial + m )^{-1} 
\not \!\! A \right]
\;,
\label{a03}
\ee
which, in momentum space representation, has the equivalent expression
\be
\Gamma (A) \;=\; \frac{1}{2} \int \frac{d^2 k}{(2 \pi)^2}
{\tilde A}_\mu (-k) {\tilde \Gamma}_{\mu\nu} (k) {\tilde A}_\nu (k)
\label{a04} \;.
\ee
Here the tilde denotes the Fourier transform of a function of $x$, 
namely, 
\ba
{\tilde A}_\mu (k) &=& \int d^2 x \, e^{-i k \cdot x} A_\mu (x)\nonumber\\ 
{\tilde \Gamma}_{\mu\nu}(k) &=& \int d^2 x \, e^{-i k\cdot x} 
\Gamma_{\mu\nu} (x) \;.
\label{a05}
\ea
Fourier transforming in (\ref{a03}), we can give a more explicit expression 
for ${\tilde \Gamma}$ as follows
$$
{\tilde \Gamma}_{\mu\nu}(k)\;=\; -e^2 \int \frac{d^2 p}{(2\pi)^2} 
{\rm tr} \left[\frac{1}{i(\not\!k+\not\!p)+m}\gamma_\mu
\frac{1}{i\not\!p+m}\gamma_\nu \right] 
$$
$$
=\; e^2 \left\{\int \frac{d^2 k}{(2\pi)^2} \frac{{\rm tr}[(\not\!k +
\not\!p)\gamma_\mu \not\!p \gamma_\nu]}{ [(k+p)^2+m^2] (p^2+m^2)}
\right.
$$
\be
\left.-m^2 \int \frac{d^2 k}{(2\pi)^2} \frac{{\rm tr} (\gamma_\mu
\gamma_\nu)}{[(k+p)^2+m^2] (p^2+m^2)} \right\} \;.
\label{a06}
\ee
Gauge invariance implies the transversality of ${\tilde \Gamma} (k)$,
which allows we to write ${\tilde \Gamma}_{\mu\nu}$ in terms of only 
one independent scalar function ${\tilde \Gamma}$
\be
{\tilde \Gamma}(k) \;=\; {\tilde \Gamma}(k^2)
(\delta_{\mu\nu}-\frac{k_\mu k_\nu}{k^2} ) \;.
\label{a07}
\ee

We follow a procedure analogous to the one followed in \cite{zinn} for
the massless case, to evaluate ${\tilde \Gamma}$ using dimensional
regularization. Thus, in $d$ dimensions, after taking the trace on both
sides of the previous tensorial equation, 
$$
(d-1) {\tilde \Gamma}(k^2)\;=\;  2 e^2 (2-d) \int \frac{d^d p}{(2\pi)^d}
\frac{(k+p)\cdot p}{[(k+p)^2+m^2] (p^2+m^2)} 
$$
\be
- 2 e^2 m^2 d \int \frac{d^d p}{(2\pi)^d} 
\frac{1}{[(k+p)^2+m^2] (p^2+m^2)} \;.
\label{a08}
\ee

Letting $d \to 2$,  the first term on the right hand side of (\ref{a08})
contributes only through its divergent part, which is identical to the 
one for the case $m=0$
\be
\lim_{d \to 2} (2-d) \int \frac{d^d p}{(2\pi)^d}
\frac{p^2}{(p^2+m^2)^2} \;=\; \frac{1}{2 \pi} \;.
\label{a09}
\ee
The second is regular at $d=2$. Adding the contribution of both terms, 
we see that 
\be
{\tilde \Gamma}(k^2) \;=\; \frac{e^2}{\pi} \;-\;
4 e^2 m^2 \int \frac{d^2 p}{(2\pi)^2}  
\frac{1}{[(k+p)^2+m^2] (p^2+m^2)} \;.
\label{a10}
\ee
The integral over $p$ is evaluated by using the standard Feynman 
parameter trick, to obtain
\newpage
\be
{\tilde \Gamma}(k^2) \;=\; \frac{e^2}{\pi} \left\{1 
- 2 \frac{m^2}{k^2} (1+\frac{4m^2}{k^2})^{-\frac{1}{2}} 
\ln \left[
\frac{(1+\frac{4m^2}{k^2})^{\frac{1}{2}} + 
1}{(1+\frac{4m^2}{k^2})^{\frac{1}{2}} - 1}               
\right] \right\}   
\label{a11}
\ee

In a low momentum expansion, the leading contributions  are
\be
{\tilde \Gamma}(k^2) \;=\; \frac{e^2}{\pi} \left\{
\frac{k^2}{6 m^2} - \frac{(k^2)^2}{90 m^4} \;.
\right\}
\label{a12}
\ee

\newpage

\vspace{3 cm}

\epsfxsize=6.in
\epsffile[100 100 600 700]{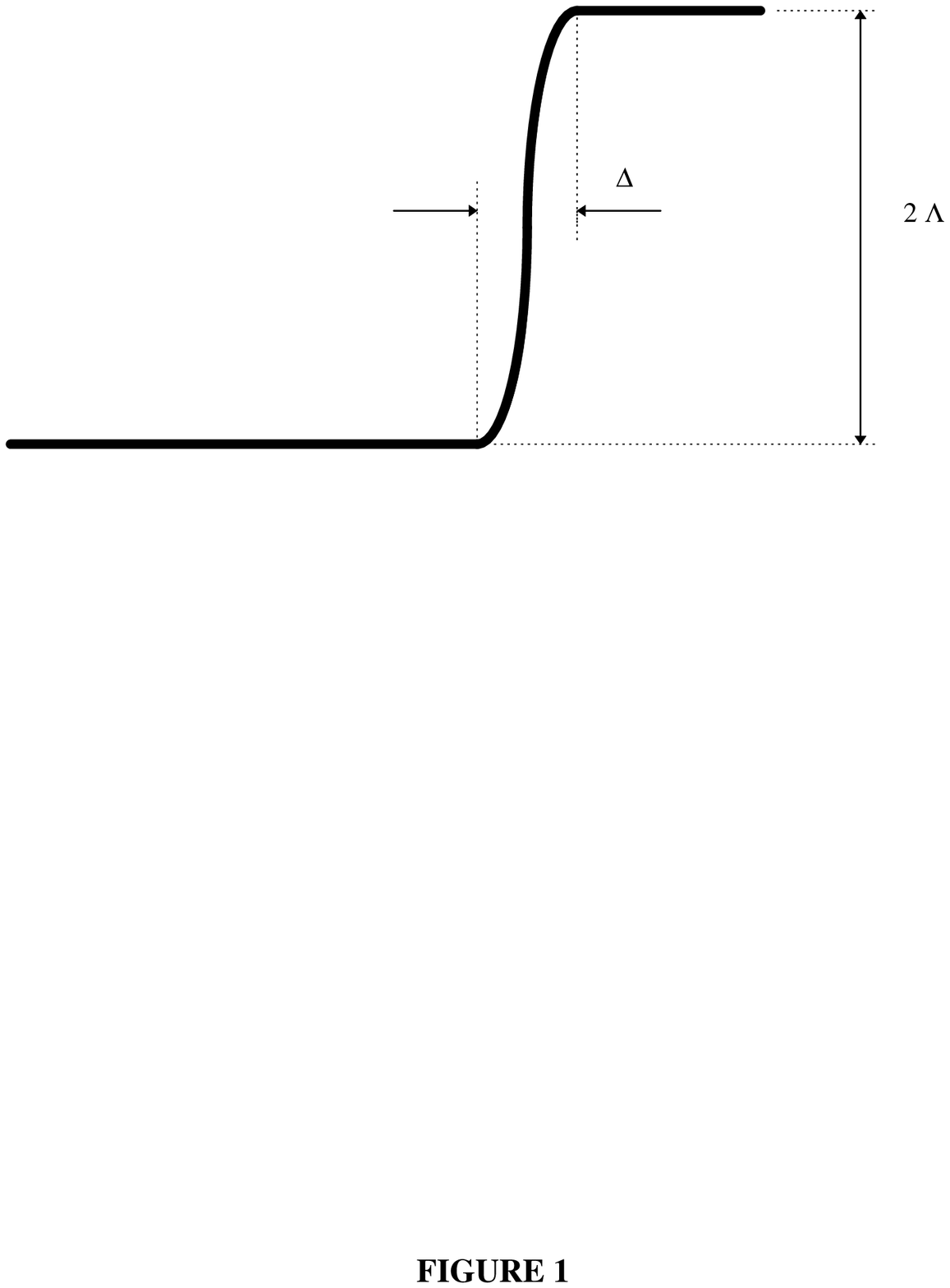}
\newpage

{\bf Figure captions}

\vspace{0.3 cm}
{\bf Figure 1} The typical mass profile considered, showing the 
definitions of the defect height $2 \Lambda$ and width $\Delta$.
\end{document}